\documentclass[twocolumn,superscriptaddress]{revtex4-2}

\usepackage[utf8]{inputenc}
\usepackage{amsmath}
\usepackage{graphicx}
\usepackage{float}
\usepackage{braket}
\usepackage{bbm} 
\usepackage{color}
\usepackage{siunitx}

\usepackage{subcaption}

\newcommand{\srr}{\hat{\sigma}^{rr}}
\newcommand{\sgr}{\hat{\sigma}^{gr}}
\newcommand{\srg}{\hat{\sigma}^{rg}}
\newcommand{\sgg}{\hat{\sigma}^{gg}}
\newcommand{\ddt}{\frac{\text{d}}{\text{d}t}}
\newcommand{\rfac}{r_\text{f}}
\newcommand{\gfac}{\Gamma_\mathrm{f}}

\newcommand{\daniel}[1]{\textcolor{black}{#1}}
\newcommand{\mfl}[1]{\textcolor{black}{#1}}

\begin{document}

\begin{abstract}
    The spread of excitations by Rydberg facilitation bears many similarities to epidemics. Such systems can be modeled with Monte-Carlo simulations of classical rate equations to great accuracy as a result of high dephasing. 
    \mfl{Motivated by experiments} we \mfl{theoretically} analyze the dynamics of a Rydberg many-body system in the facilitation regime in the limits of high and low temperatures.
    \daniel{In} the high-temperature limit a homogeneous mean-field \daniel{behavior} is recovered\daniel{, while} characteristic effects of heterogeneity can be seen in a frozen gas. At \daniel{high} temperatures the system displays an absorbing-state phase transition and, in the presence of an additional loss channel, self-organized criticality. In a frozen or low-temperature gas, excitations are constrained to a network resembling an Erdős–Rényi graph.  We show that the absorbing-state phase transition is replaced with an extended Griffiths phase, which we accurately describe by a susceptible-infected-susceptible model on the Erdős–Rényi network taking into account Rydberg blockade.
\end{abstract}

\title{Griffiths Phase in a Facilitated Rydberg Gas at Low Temperatures}

\author{Daniel Brady}
\affiliation{Department of Physics and Research Center OPTIMAS, University of Kaiserslautern, D-67663 Kaiserslautern, Germany}
\author{Jana Bender}
\affiliation{Department of Physics and Research Center OPTIMAS, University of Kaiserslautern, D-67663 Kaiserslautern, Germany}
\author{Patrick Mischke}
\affiliation{Department of Physics and Research Center OPTIMAS, University of Kaiserslautern, D-67663 Kaiserslautern, Germany}
\affiliation{Max Planck Graduate Center with Johannes Gutenberg University Mainz (MPGC), D-55128 Mainz, Germany}
\author{\mfl{Simon Ohler}}
\affiliation{Department of Physics and Research Center OPTIMAS, University of Kaiserslautern, D-67663 Kaiserslautern, Germany}
\author{Thomas Niederprüm}
\affiliation{Department of Physics and Research Center OPTIMAS, University of Kaiserslautern, D-67663 Kaiserslautern, Germany}
\author{Herwig Ott}
\affiliation{Department of Physics and Research Center OPTIMAS, University of Kaiserslautern, D-67663 Kaiserslautern, Germany}
\author{Michael Fleischhauer}
\affiliation{Department of Physics and Research Center OPTIMAS, University of Kaiserslautern, D-67663 Kaiserslautern, Germany}

\date{\today}

\maketitle

\section{Introduction}

\begin{figure}[H]
  \centering
  \includegraphics[width=\columnwidth]{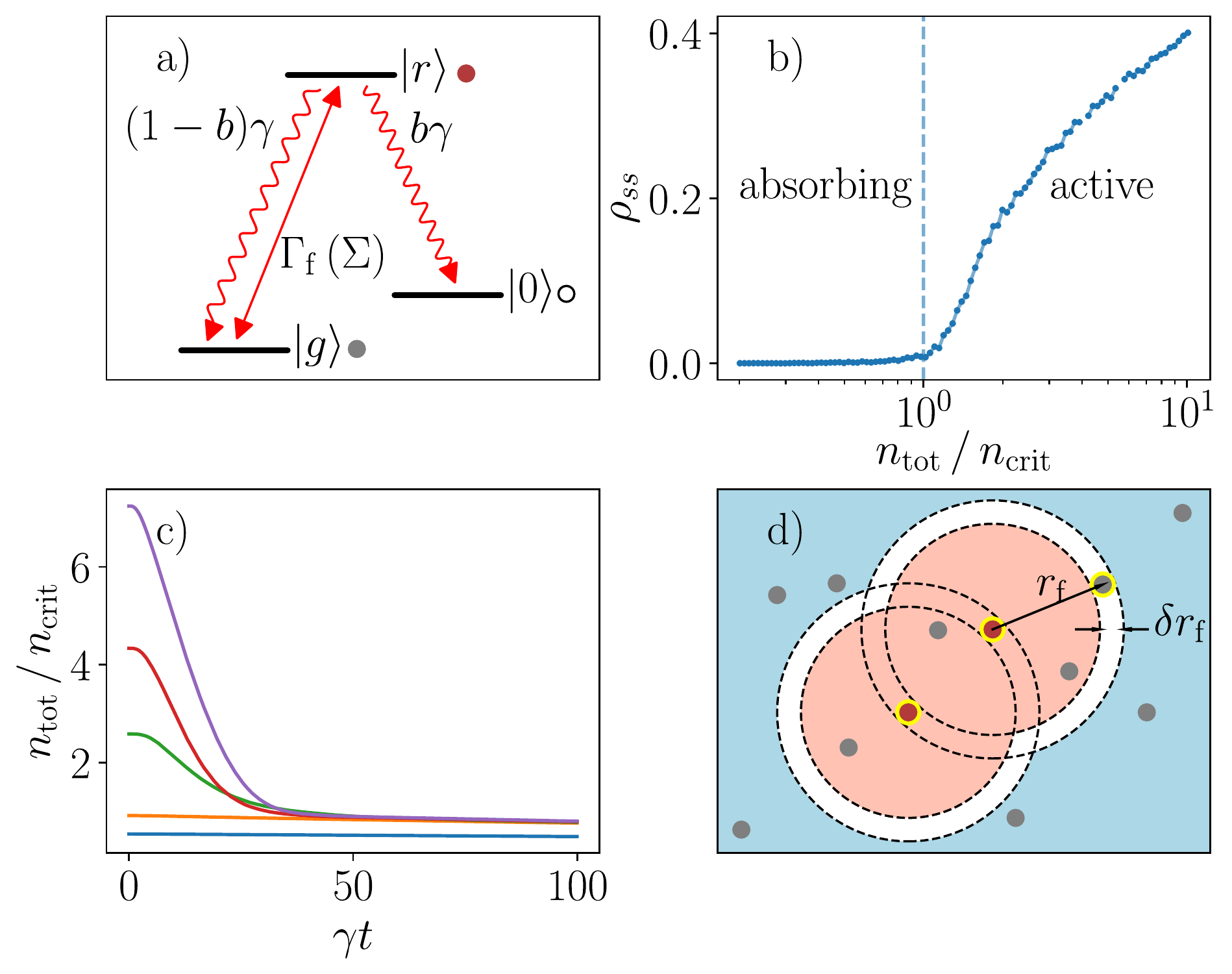}
  \caption{(a): \daniel{Laser field couples ground $\ket{g}$ and Rydberg $\ket{r}$ states} \mfl{resulting in a} \daniel{transition rate $\gfac(\Sigma)$ (see eq.~\eqref{eq:gamma_fac}). The Rydberg state can spontaneously decay into $\ket{g}$ or an inert state $\ket{0}$.} (b): Steady state Rydberg density depending on total \mfl{active} density (\mfl{i.e. in states $\vert g\rangle$ and $\vert r\rangle$}) for ${b=0}$ from Monte-Carlo simulations. (c): Total \mfl{active} density \mfl{$n_\textrm{tot}$}  over time from Monte-Carlo simulations for ${b > 0}$ showing self-organization of the system to the critical density $n_\textrm{crit}$, if the initial density 
  is larger. (d): Schematic of facilitation shell (white), atoms (grey) in the red area are subject to Rydberg blockade and atoms in the blue area only weakly interact with the Rydberg atoms (red).
}
  \label{fig:intro}
\end{figure}

Rydberg atoms have gained a lot of attention in recent years due to their strong interactions over large distances \cite{gallagher2006rydberg}. This, paired with their long lifetimes in the order of milliseconds, creates a platform to explore quantum many-body physics of strongly interacting spin systems \cite{weimer2010rydberg,schauss2012observation,Bernien2017,Browaeys2020,Surace2020,scholl2021quantum,Leseleuc2019,Semeghini2021} and to implement key elements for quantum information processing \cite{PhysRevLett.85.2208,PhysRevLett.87.037901,gaetan2009observation,urban2009observation,saffman2010quantum}. Moreover,
optically driven Rydberg atoms (see Fig.~\ref{fig:intro}a) can be used to investigate many-body dynamics of spin systems in inherently dissipative environments \cite{PhysRevA.98.022109,PhysRevLett.112.013002,malossi2014full,urvoy2015strongly,letscher2017bistability}, as the laser excitation into high lying Rydberg states is often accompanied by strong dephasing. The latter includes important dynamical phenomena such as an absorbing-state phase transition (see Fig.~\ref{fig:intro}b), one of the simplest classical non-equilibrium phase transitions displaying critical behavior and universality \cite{doi:10.1080/00018730050198152,henkel2008non}.

Absorbing-state phase transitions 
are of general interest as they occur in many phenomena outside of physics such as population dynamics, epidemics, or the spreading of information in social media \cite{grassberger1983critical,kuhr2011range,bonachela2012patchiness,xie2021detecting}. Systems with this phase transition 
\mfl{are believed} fall into the  universality class of directed percolation (DP) \cite{doi:10.1080/00018730050198152}. The unambiguous experimental observation of DP 
universal behavior is however challenging and has only been achieved in few systems in recent years \cite{hinrichsen2000possible,rupp2003critical,takeuchi2007directed,takeuchi2009experimental,lemoult2016directed,kohl2016directed}. 
More recently\daniel{,} experimental signatures of such a transition have been reported in optically driven Rydberg gases \cite{gutierrez2017experimental}. 

In Rydberg systems,
dissipation can give rise to another important dynamical phenomenon: self-organized criticality (SOC) \cite{bak1988self, bak2013nature}, which is believed to be a cause for the abundance of scale-invariance in nature \cite{sornette1989self,rhodes1996power,malamud1998forest,hesse2014self}. An SOC system dynamically evolves to the critical point of a phase transition by itself due to dissipation \daniel{and} without the need for parameter fine tuning (see Fig.~\ref{fig:intro}c). Since the dissipation is strongly reduced once the critical point is reached, further evolution into the absorbing phase happens on much longer time scales.
Recent experiments on Rydberg facilitation have shown some evidence of SOC through the use of ionization, or a decay into an auxiliary \daniel{inert/}"dead" state as a loss mechanism (see Fig.~\ref{fig:intro}a) \daniel{\cite{helmrich2020} (see \cite{Ding_etal} for related experiments)}.

However, the directed percolation transition is known to be susceptible to disorder \cite{PhysRevE.72.036126} and more recent experiments on Rydberg facilitation in a trapped ultra-cold gas of atoms gave indications for an emergent heterogeneity in the system \cite{natcom_Griffiths}. In such a heterogeneous system, the critical point of the absorbing-state phase transition is replaced by an intermediate extended Griffiths phase. Griffiths phases are characterized by generic scale invariance and the lack of universal behavior. This is in contrast to an absorbing state phase transition where scale invariance is only expected at the critical point.
As a result (e.g. in the Rydberg gas), one expects a power law decay in active density over time with continuously varying exponents depending on the driving strength \cite{PhysRevLett.23.17}.

In \cite{natcom_Griffiths}, it was experimentally shown that a Rydberg system in the facilitation regime produces signatures of such a Griffiths phase for short times compared to the lifetime of the Rydberg state. A power-law decay in Rydberg density over time was observed with the decay exponents varying with driving strength and a phenomenological susceptible-infected-susceptible (SIS) network model was put forward to describe the observations. The model included a fitting function for the node weights of the network depending on the excitation rate $\kappa$.
The interpretation being, that in the network model heterogeneity originates from a velocity selective excitation mechanism, where only atoms with relative velocities smaller than the Landau-Zener velocity $v_\textrm{LZ}(\kappa)$ could participate in facilitation dynamics. Above this velocity all further excitations are exponentially suppressed.

\daniel{In the present paper, we \mfl{present experimental} indications for generic scale-invariance and strong \mfl{theoretical} indications for a Griffiths phase in a Rydberg facilitation gas by Monte-Carlo simulations. }

In the experiment we continuously monitor the number of Rydberg excitations in a trapped ultra-cold gas of $^{87}$Rb atoms. We show that the size distribution of the Rydberg excitation number follows a power-law distribution, i.e. shows a scale free behavior, over an extended parameter regime, which is a key characteristic of a Griffiths phase.
 
In order to understand and to quantitatively describe the emergence of the Griffiths phase, we theoretically analyze two limiting cases: (i) a frozen gas and (ii) a gas with high temperature.  While we recover a direct absorbing-state phase transition in the high-temperature limit with no signs of a velocity induced heterogeneity, we can identify a Griffiths phase in the \emph{frozen gas limit} as a result of the finite paths along which facilitated excitations can spread. We give a quantitative analysis of the factors contributing to the emergence of a Griffiths phase and provide an estimate for the characteristic exponents of the power-law decay of Rydberg activity in this phase.

The facilitation of Rydberg excitations in a gas of optically driven atoms can be microscopically described by a Lindblad master equation \cite{lindblad} for the density matrix $\hat{\rho}$, which takes the form
\begin{align}
    \label{eq:master_equation}
    \ddt \hat{\rho} = i [\hat{\rho}, \hat{\mathcal{H}}] 
    + \sum_l \hat{L}_l \hat{\rho} \hat{L}_l^\dagger
    - \frac{1}{2} \{ \hat{L}_l^\dagger \hat{L}_l, \hat{\rho} \}.
\end{align}
Here, the atom-light interaction Hamiltonian $\hat{\mathcal{H}}$ is given by
\begin{align}
    \label{eq:hamiltonian}
    \hat{\mathcal{H}} &= \sum_i
        \Big[
            \Omega (\sgr_i + \srg_i)
            +
            \Big(
                \sum_{j \neq i} \frac{c_6}{r_{ij}^6} \srr_j - \Delta
            \Big)
            \srr_i
        \Big],
\end{align}
where $\hat\sigma_j^{\mu\nu} = \vert \mu\rangle_{jj}\langle \nu\vert$
is the transition operator between states $\vert \nu\rangle$ and $\vert \mu\rangle$ of the $j$th atom. The strength of the laser driving shifted from the ground-Rydberg resonance frequency
by the detuning $\Delta$ is described by the Rabi-frequency $\Omega$, and there is a van der Waals interaction 
proportional to $c_6/r_{ij}^6$, with $r_{ij}=\vert \vec r_i-\vec r_j\vert$ being the distance between atoms $i$ and $j$.
Dissipative processes are taken into account by the Lindblad jump operators ${\hat{L}_1^{(i)} = \sqrt{(1-b)\gamma} \sgr_i}$, ${\hat{L}_2^{(i)} = \sqrt{b\gamma} \hat \sigma_i^{r0}}$ describing spontaneous decay of the Rydberg state into the ground state $\ket{g}$ and the inert state $\ket{0}$, with the branching parameter $b$. \daniel{Finally, dephasing, attributed to laser phase noise and Doppler broadening \cite{helmrich2020} as well as the spread of the atomic wave packet over the van-der-Waals potential \cite{motional_dephasing_lesanovsky}, is described by ${\hat{L}_\perp^{(i)} = \sqrt{\gamma_\perp} \srr_i}$.}

The strong van der Waals interaction of a Rydberg atom shifts energy levels of surrounding atoms significantly up to distances of multiple \SI{}{\micro\meter}. When the atoms are resonantly coupled to a laser field, this will block further excitations into Rydberg states from occurring for all atoms within a finite distance, a phenomenon known as Rydberg blockade \cite{lukin2001dipole}. On the other hand, if the laser excitation is strongly detuned, the excitation of isolated atoms is suppressed while atoms close to the facilitation distance $\rfac \equiv \sqrt[6]{\frac{c_6}{\Delta}}$ are shifted into resonance (Fig.~\ref{fig:intro}d) and are excited with a greatly increased rate. This process, termed Rydberg facilitation, leads to a cascade of excitations quickly spreading through the system following a single (off-resonant) excitation  \cite{PhysRevLett.98.023002, PhysRevLett.104.013001}. It is important to note that Rydberg blockade still occurs in this regime. The excitation of atoms with distances ${r < \rfac}$ \daniel{is} greatly suppressed (red zone in Fig.~\ref{fig:intro}d).

\section{Experimental observation of scale-free behavior in a driven Rydberg gas}

To experimentally \mfl{test} \daniel{scale-invariance}, we investigate the excitation density in a trapped gas of $^{87}$Rb atoms. To this end, we prepare a sample containing \num{150e3} atoms at a temperature of \SI{1}{\micro \kelvin} in a crossed optical dipole trap. 
The sample has a density on the order of $10^{12}/$\si{\centi \meter \cubed}.
From the $5\mathrm{S}_{1/2}$ ground state, a UV laser at \SI{297}{\nano \meter} continuously couples to the $40\mathrm{P}_{3/2}$ Rydberg state with a detuning of +\SI{40}{\mega \hertz} and a resonant Rabi frequency of $2\pi\times\SI{100}{\kilo  \hertz}$.
The temperature of the gas corresponds to a most probable speed $\hat{v} = 0.7 \, \rfac \gamma$ with the facilitation radius $\rfac $ and decay rate $\gamma $.

\begin{figure}[H]
    \centering
    \includegraphics[width=\columnwidth]{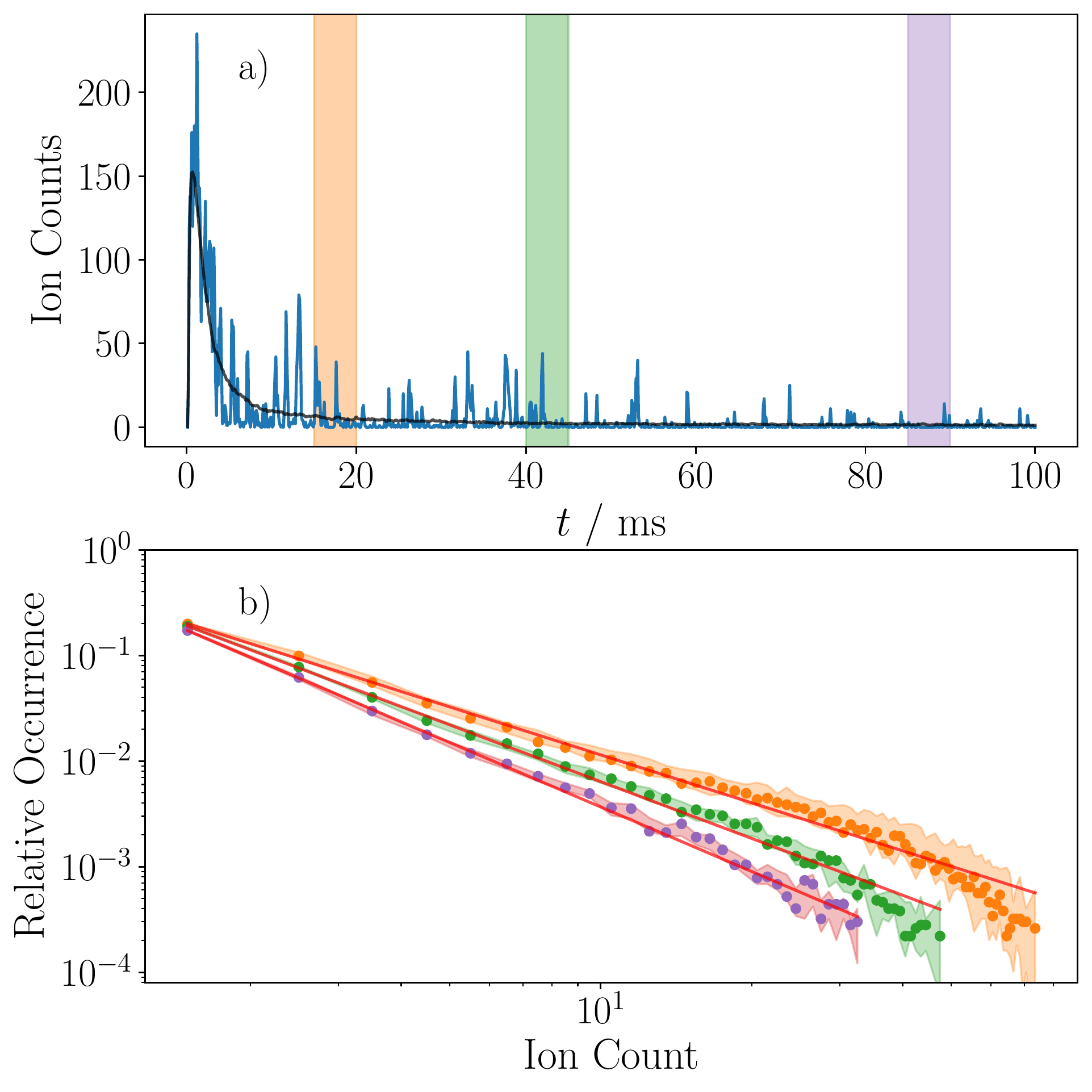}
    \caption{(a): Ion signal per \SI{10}{\micro \second} time interval for a single experiment run (blue line) and average over \num{1000} experimental runs (black line). The ground state is off-resonantly coupled to the Rydberg state for \SI{100}{\milli \second}. During the measurement, the density continuously decreases because of the intrinsic ionization of Rydberg atoms. In the first few milliseconds, the system is in the active phase displaying continuously high activity.  Afterwards, the dynamics is dominated by isolated avalanches. The colored areas indicate the time segments evaluated in (b). (b): Experimentally found distribution of ion counts for different sample densities averaged over \num{1000} experimental runs. We choose exemplary \SI{5}{\milli \second}-long time segments at \SI{15}{\milli \second} (orange), \SI{40}{\milli \second} (green) and \SI{85}{\milli \second} (violet) corresponding to three densities. The distributions show power-law behavior (fitted in red), albeit with distinct exponents (\num{-1.51}, \num{-1.79}, and \num{-2.03} respectively). \daniel{The shaded region \mfl{characterizes the uncertainty in the measurements. It} represents the maximum/minumum relative occurrence when shifting the evaluation windows by $\pm 5$~ms.}}
    \label{fig:experiment}
\end{figure}

Atoms in the $40\mathrm{P}_{3/2}$ state are ionized because of multiple intrinsic processes \cite{schlagmuller_ultracold_2016, niederprum_giant_2015}, which we use to continuously monitor the excitation number. To this end
we guide the resulting ions to a detector using a small electric field.
This yields a time-resolved signal proportional to the number of Rydberg excitations in the sample (Fig.~\ref{fig:experiment}a).

At the beginning of the continuous laser exposure, which lasts \SI{100}{\milli \second}, there are no excitations in the sample. As soon as the first off-resonant excitation is created, activity spreads through the system via facilitation, setting it up in the active phase. Due to the continuous atom loss caused by the ionization of excited atoms, the sample density decreases, reducing the effective driving strength. The sample thus approaches the phase transition.

We divide the ion signal in segments of \SI{5}{\milli\second} to account for the temporally varying effective driving.
For each of these segments, we analyze the ion count distribution in \SI{10}{\micro \second} bins and average over \num{1000} experimental runs.
After about \SI{10}{\milli \second} the average activity has dropped more than an order of magnitude compared to its maximum value, while in individual runs it is dominated by avalanches. Therefore, we assume that at this time the sample is leaving the active phase.

Our measurement data shows persistent power-law behavior in the distribution of avalanche sizes over a wide range of densities (Fig.~\ref{fig:experiment}b). Power laws are a clear signature of scale invariance, which is expected only at the critical point of an absorbing-state phase transition or in a Griffiths phase characterizing a heterogeneous system.
The extracted exponent of the power-law distribution is not fixed but varies with the density, strongly indicating non-universal behavior. 
\mfl{While these observations are not an experimental proof of heterogeneity we use them as motivation to
theoretically investigate possible origins of heterogeneity and a related Griffiths phase in the system.}
\ 

\section{Microscopic Model of Rydberg facilitation}

\daniel{After having shown indications of scale-invariant behavior in the Rydberg facilitation gas, \mfl{however with varying exponents in the} experiments, we now turn to a theoretical modelling} \mfl{of the microscopic dynamics.} 

In the limit of large dephasing, the dynamics of a many-body Rydberg gas are effectively governed by classical rate equations \cite{Levi_2016}. As such, we will simulate a gas of atoms governed by \eqref{eq:master_equation} using classical Monte-Carlo simulations of a set of rate equations derived from \eqref{eq:master_equation} in the limit of large dephasing. After adiabatic elimination of coherences, eq. \eqref{eq:master_equation} reduce to classical rate equations between ground, Rydberg, and inert states (see Fig.~\ref{fig:intro}a), with the stimulated rate $\gfac(\Sigma)$ given as
\begin{align}
    \label{eq:gamma_fac}
    \gfac(\Sigma) = \frac{2 \Omega^2 \gamma_\perp}{\gamma_\perp^2 + \Delta^2
    \big(
    \sum_{\substack{j \neq i \\ j \in \Sigma}} \frac{\rfac^6}{r_{ij}^6} - 1
    \big)^2},
\end{align}
where $\Sigma$ is the set of indices of Rydberg-excited atoms. \daniel{In order to ensure numerical stability in the simulation, the singularity of the potential in eq.~\eqref{eq:gamma_fac} is truncated at a cutoff value.} 

\daniel{Using eq.~\eqref{eq:gamma_fac} we can formulate a set of classical rate equations for the probability of the $i$th atom being in the Rydberg state $P_\mathrm{r}^{(i)}$ or the ground state $P_\mathrm{g}^{(i)}$ as}
\begin{subequations}
    \label{eq:rate_equations}
    \begin{align}
        \label{eq:rate_equations_ryd}
        \ddt P_\mathrm{r}^{(i)} &= \gfac(\Sigma) P_\mathrm{g}^{(i)} - (\gfac(\Sigma) + \gamma) P_\mathrm{r}^{(i)}
        \\
        \label{eq:rate_equations_grd}
        \ddt P_\mathrm{g}^{(i)} &= (\gfac(\Sigma) + (1- b)\gamma)  P_\mathrm{r}^{(i)} - \gfac(\Sigma) P_\mathrm{g}^{(i)}.
    \end{align}
\end{subequations}

If no other Rydberg atom exists in the gas or their distance is much larger than $\rfac$, $\gfac(\Sigma)$ reduces to the off-resonant excitation rate of an isolated atom
\begin{align}
    \tau = \frac{2 \Omega^2 \gamma_\perp}{\gamma_\perp^2 + \Delta^2}.
\end{align}

\daniel{As a result of the broadening of the ground-Rydberg transition, given by the dephasing rate $\gamma_\perp$, facilitation can occur in a smeared out region around the facilitation distance $\rfac$, given by}
\begin{align}
    \delta \rfac = \frac{\gamma_\perp}{2 \Delta} \rfac.
\end{align}
\daniel{Therefore, each Rydberg atom spans a facilitation shell around it at the radius $\rfac$ and with the width ${\delta \rfac}$ (white disks in Fig.~\ref{fig:intro}d).} Inside this shell, the stimulated rate takes its maximal value ${\gfac = \frac{2\Omega^2}{\gamma_\perp}}$ referred to as the facilitation rate. Relevant for 
later mappings to epidemic models is this rate integrated over the volume $V_s$ of the facilitation shell given by
\begin{align}
    \kappa = \gfac \, V_s.
\end{align}

The relevant quantities of interest here are the coarse grained Rydberg density (in a small volume $\Delta V$)
\begin{align}
    \label{eq:coarse_grain_rho}
    \rho(\vec{r},t) = \frac{1}{\Delta V}\sum_{i:\vec{r}_i\in \Delta V} \langle \srr_i\rangle,
\end{align}
and \daniel{the} total \mfl{active} density of ground\mfl{-state} and Rydberg atoms
\begin{align}
    \label{eq:coarse_grain_n}
    n(\vec{r},t) = \frac{1}{\Delta V}\sum_{i:\vec{r}_i\in \Delta V} \Bigl(\langle \srr_i\rangle + \langle \sgg_i\rangle\Bigr).
\end{align}
\daniel{In the following, $n$ will be referred to as the total  density of the gas, \mfl{for} \daniel{simplicity}. As atoms in the state $\ket{0}$ do not participate in the dynamics of the system (see Fig.~\ref{fig:intro}a), a decay into this state corresponds to a reduction of the total density, i.e. atom loss.}

With this, $n \kappa$ corresponds to the rate with which excitations spread through the cloud.

The gas is simulated in a cube with size $L^3$ and periodic boundary conditions, typically ${L=7 \, \rfac}$. Atom positions are chosen randomly and velocities are sampled from the Maxwell-Boltzmann distribution with the temperature parameter $\hat{v}$, corresponding to the most probable atom velocity in the gas.
After choosing a fixed time-step (${dt=1/400 \; \gamma}$), the time evolution of the system is given by a fixed time step Monte-Carlo (ftsMC) algorithm \cite{RUIZBARLETT20095740}. We choose an ftsMC algorithm as opposed to a kinetic Monte-Carlo algorithm \cite{Chotia_2008} as atomic movement, paired with long-range interactions leads to quickly changing transitional rates in the system.

In Ref.~\cite{helmrich2020} Langevin equations have been derived to macroscopically describe the density of Rydberg atoms $\rho$ and the total density
$n$ in the system. \daniel{As is shown in \cite{helmrich2020}, the homogeneous mean-field solution, in which diffusion terms are neglected, is sufficient to model the system. These equations then take the form}
\begin{subequations}
    \begin{align}
        \label{eq:langevin_rho}
        \ddt \rho &= 
        -\kappa (2 \rho^2 - \rho n)
        -\gamma \rho
        -\tau (2\rho - n) + \xi,
        \\
        \label{eq:langevin_n}
         n &= n_0 - b\gamma \int_0^{t} \text{d}t^\prime \: \rho(t^\prime),
    \end{align}
    \label{eq:langevin_diehl}
\end{subequations}
with the off-resonant excitation rate $\tau$, and a noise term $\xi$. The parameter $b$ \daniel{characterizes the percentage of Rydberg atoms which spontaneously decay into the dead state $\ket{0}$ (see Fig.~\ref{fig:intro}a). As mentioned above, atoms that decay into this state are effectively removed from the system.}

\daniel{Assuming a gas with a heterogeneous density, diffusion results in a stabilization of the critical point over long times. For details pertaining to this see \cite{klocke2021hydrodynamic}.}

\begin{figure}[H]
  \centering
  \includegraphics[width=\columnwidth]{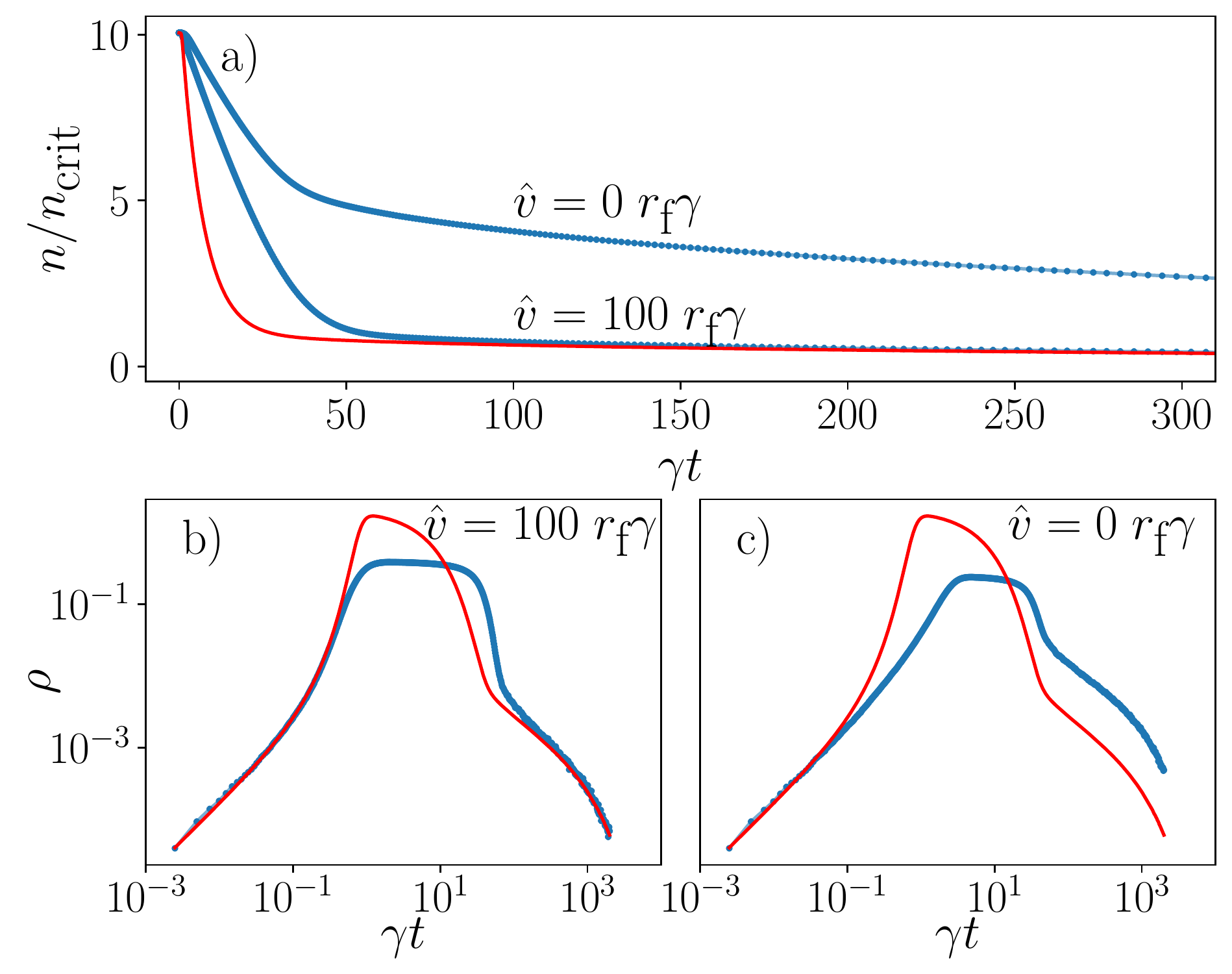}
  \caption{(a): Sum of Rydberg and ground state atom densities over time from Monte-Carlo simulations (blue dots) for ${\hat{v}=0 \; \rfac \gamma}$ and ${\hat{v} = 100 \; \rfac \gamma}$ compared with the prediction from the Langevin equations \eqref{eq:langevin_diehl} from \cite{helmrich2020} (red line). (b): Rydberg density for ${\hat{v} = 100 \; \rfac \gamma}$. (c): Rydberg density for ${\hat{v} = 0 \; \rfac \gamma}$. For all plots we use the parameters: ${n_0 = 4 \; \rfac^{-3}}; \; {\Omega / \gamma = 20}; \; {\Delta / \gamma = 2000};$ ${\gamma_\perp / \gamma = 20}; \; {b=0.3}; \; {L / \rfac = 7}$.
  }
  \label{fig:total_densities}
\end{figure}

In the absence of decay into $\vert 0\rangle$, i.e. for ${b=0}$, and in the absence of an off-resonant excitation, i.e. $\tau =0$, the dynamics described by eqs.~\eqref{eq:langevin_diehl}, feature an absorbing-state phase transition at the critical atom density
\begin{align}
    \label{eq:n_crit}
    n_\textrm{crit} = \frac{\gamma}{\kappa},
\end{align}
when the facilitation rate is fixed, or alternatively, at the critical facilitation rate $\kappa_\textrm{crit}= \gamma/n_0$ for fixed density.
Below the critical point, any initially existing excitations in the system will eventually decay and the steady state of the system is one where all atoms are in the ground state (absorbing phase).
Above the critical point, any arbitrarily small number of excitations initially present in the system will facilitate further excitations cascading through the system until a steady state with finite excitation density ${\rho(t \to \infty) > 0}$ is reached (active phase). 

Off-resonant excitations, with the rate $\tau$, will seed an excitation cascade in the active phase; whereas, in the absorbing phase, they cause fluctuations in the excitation number. As a result, the true absorbing state ${\rho = 0}$ can only be approximately reached experimentally through a large separation of the off-resonant and facilitation time-scales, suppressing off-resonant excitations on the experimentally relevant facilitation time-scales.

Finally, the (slow) decay into a dead state $\vert 0\rangle$ with rate $b\gamma$ is responsible for the 
self-organized approach to the critical point when starting in the active phase as indicated in Fig.~\ref{fig:intro}c. Starting at an initial density $n_0$ above the critical value $n_\textrm{crit}$, i.e. in the active phase, the large number of atoms in the Rydberg state causes a fast loss of atoms into the dead state. As a consequence, the total density of atoms $n$ effectively participating in the facilitation process, i.e. atoms in states $\vert g\rangle$ and  $\vert r\rangle$ decreases quickly and approaches the critical value. This loss continues at the critical density and drives the system further into the absorbing state. However, this happens on a much slower time-scale, as fewer Rydberg excitations are present at the critical point.

In Fig.~\ref{fig:total_densities} we have plotted the time evolution of the total density $n$, initially ten times higher than the critical density $n_\textrm{crit}$, and the Rydberg density $\rho$ for a frozen gas as well as a high-temperature gas with otherwise identical conditions, obtained from Monte-Carlo simulations. Here, all atoms in the system are initially in the ground state until one atom is off-resonantly excited to the Rydberg state.
For comparison we also show the solution of the mean-field Langevin equations \eqref{eq:langevin_diehl}, which 
capture the long-time SOC dynamics of the high-temperature gas, but fail to describe the frozen gas outside of very short times (see Figs.~\ref{fig:total_densities}b and c). The discrepancy in the peak values of $\rho$
can be attributed to Rydberg blockade, which truncates the maximum number of Rydberg excitations simultaneously present in the gas.

Qualitatively, the Rydberg density in the frozen gas displays a similar time dynamic to that of the high temperature gas, albeit with substantial quantitative differences in the long-time limit. We will show that in the low temperature regime of the Rydberg gas the absorbing state phase transition is  replaced with an extended Griffiths phase, whose characteristic features become visible when off-resonant excitations and decay into state $\ket{0}$ are negligible.

It is important to note that for ${b > 0}$ the decay into $\ket{0}$ dominates the dynamics at times ${t > 1 / b\gamma}$. Thus in order to experimentally observe a Griffiths phase by monitoring the long-time dynamics with this system, ionization and loss of Rydberg atoms (manifested in the parameter $b$) must be reduced as much as possible.

\daniel{
In Ref.~\cite{natcom_Griffiths} it was argued that a Rydberg atom moving at an average velocity larger than the Landau-Zener velocity ${v_\textrm{LZ}= 2\pi^2 \Omega^2 \rfac / (3\Delta)}$,
effectively decouples from the excitation cascade. As a result, it was argued that this system features an emerging heterogeneity at high temperatures. Considering the two limiting cases of a frozen gas and a high temperature gas we argue that the Griffiths phase, which originates from spatial inhomogeneity, disappears when the atoms average velocity is increased above a certain limit, resulting in a direct absorbing-state phase transition. 
}

\daniel{
A quantitative discussion of the crossover between a frozen system with an extended Griffiths phase  and a high-temperature gas with a direct absorbing-state phase transition is beyond the scope of the present paper 
and is subject to future work. Instead we will focus on the quantitative description of the
facilitation dynamics in a low-temperature or frozen gas.
}

\section{Network Structure of Facilitation Paths in a Frozen Gas}

The emergence of a Griffiths phase 
results from facilitation events being constrained to a network structure. In the limit of a frozen gas, atoms have random but fixed positions. If we regard the system at the time scale of facilitated excitations, off-resonant excitations can be neglected. Therefore, the dynamics are described by the facilitated spreading of Rydberg excitations, which is only possible if atomic distances are approximately $\rfac$. As a result, we can regard the structure of atom positions and the paths along which excitations can spread as a random graph \daniel{with edges where atoms have the distance ${r \in [\rfac - \frac{\delta \rfac}{2}, \rfac + \frac{\delta \rfac}{2}]}$.}

Assuming a uniform distribution of atom positions in the gas, the probability that a randomly selected atom has $k$ atoms in its facilitation shell (see Fig.~\ref{fig:intro}\daniel{d}), meaning the atom is of degree $k$, is given by the Poissonian distribution
\begin{align}
    \label{eq:poissonian}
    P(k) = \frac{(n V_s)^k}{k!} \exp{(-n V_s)}.
\end{align}

As the degree distribution is Poissonian we can map this problem to a random Erdős–Rényi (ER) network \cite{erdHos1960evolution}. In contrast, the network structure of atoms trapped by an optical lattice or tweezer array would be given by a regular lattice network.

\begin{figure}[H]
    \centering
    \includegraphics[width=\columnwidth]{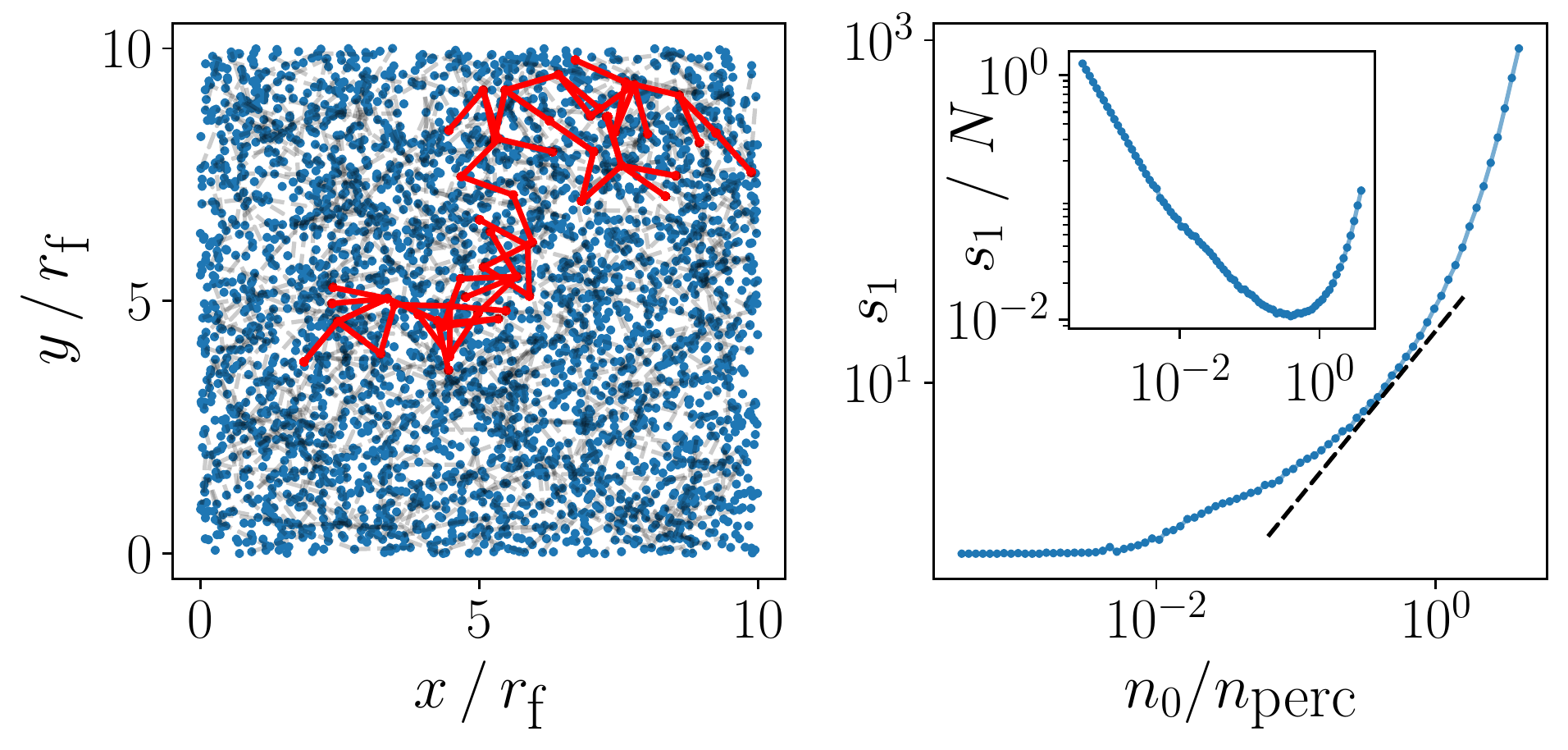}
    \caption{Schematic frozen gas atom positions (blue dots) for a 2D system with ${L=10 \; \rfac}$. Network clusters connecting atoms which have distances $r_{ij} \in [\rfac - \frac{\delta \rfac}{2}, \rfac + \frac{\delta \rfac}{2}]$ (grey dashed lines), and the largest connected cluster of these (red lines) for ${n_0 = n_\text{perc}}$ are shown (left). Size of largest connected cluster (LCC) $s_1$ depending on the density from Monte-Carlo samples in a 3D cube with ${L=7 \; \rfac}$ (right). The black dashed line corresponds to a power-law with exponent ${\nu = 1}$. And LCC divided by number of atoms $N$ depending on density (inset).}
    \label{fig:lcc}
\end{figure}

Of particular interest in random graph theory is the question if a system percolates. In a percolating system, the probability $p$ that a bond between two randomly selected atoms exists is high enough, such that a path exists which runs through the entire system, i.e. there almost surely exists a single cluster \daniel{(i.e. a single connected set of vertices)} with its size in the order of the system size. If, however, the connectivity is below a critical threshold for bond connectivity $p < p_c$, the system is composed of many small, disconnected clusters \cite{erdHos1960evolution, LI20211}. For $p = p_c$ the percolation transition occurs. A 2D network with ${p=p_c}$ from Monte-Carlo sampling is illustrated in Fig.~\ref{fig:lcc}. 

If $N$ is the number of atoms and $s_1(N)$ is the size of the largest connected cluster (LCC), then the system percolates if
%
 ${   \lim_{N\to \infty} {s_1(N)}/{N} > 0}$.
%
For an ER network the percolation transition occurs when the average network degree is $\langle k \rangle = 1$ \cite{LI20211, sayama2015introduction}. Using eq.~\eqref{eq:poissonian}, the density at which the percolation transition occurs is therefore
\begin{align}
    \label{eq:n_perc}
    n_\textrm{perc} = \frac{1}{V_s}.
\end{align}

This density is a factor $\gfac / \gamma$ larger than the critical density $n_\text{crit}$ of the absorbing state phase transition, given by eq.~\eqref{eq:n_crit}. We can verify that eq.~\eqref{eq:n_perc} corresponds to the correct percolation density by calculating the size of the LCC $s_1$ depending on the density of the gas (Fig.~\ref{fig:lcc}). In the thermodynamic limit, ${s_1 / N = 0}$ for all densities ${n < n_\textrm{perc}}$. As numeric simulations are restricted to a finite system size however, we instead consider the percolation transition to occur when $s_1$ grows faster than linear with the density $n$ (the black dashed line in Fig.~\ref{fig:lcc} corresponds to linear growth).

Of relevance for the Griffiths phase is the size distribution of clusters in the network. Using Monte-Carlo simulations we can verify that the lengths of clusters follow a geometric distribution ${P(s) \sim \text{e}^{-cs}}$ under the assumption that clusters are made of linear chains of $s$ atoms. This assumption holds true for small cluster sizes and an average network degree ${\langle k \rangle \ll 1}$.

\begin{figure}[htb]
    \centering
    \includegraphics[width=0.9\linewidth]{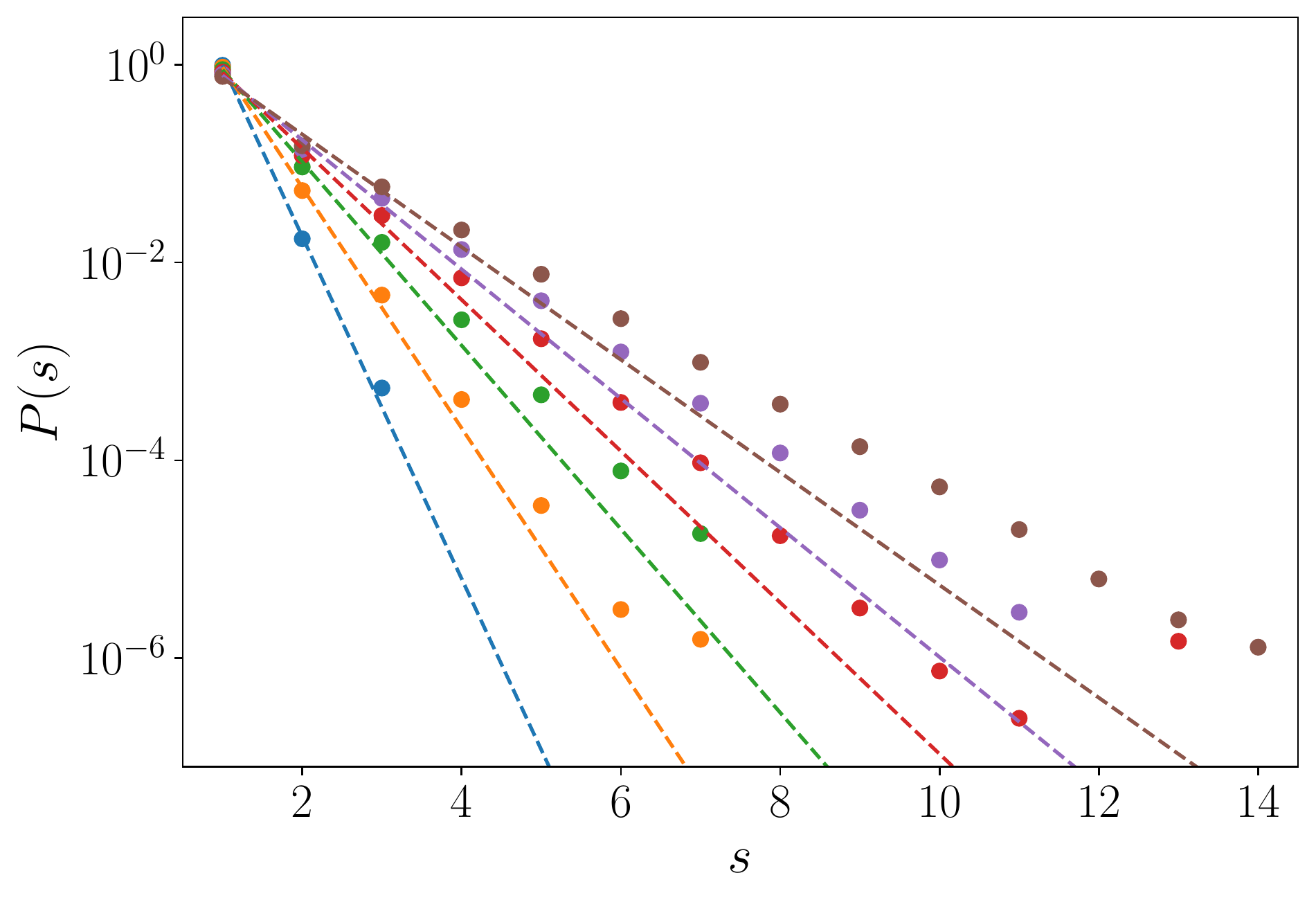}
    \caption{Network cluster size probability distribution from 3D Monte-Carlo samples (dots) and eq.~\eqref{eq:cluster_sizes} (dashed lines). From left to right for the densities ${n_0 / n_\text{perc} = [0.019, 0.063 , 0.125  , 0.188 , 0.250   , 0.313 ]}$.}
    \label{fig:cluster_sizes}
\end{figure}

We can then approximate the decay constant $c$, with $p_0$ being the probability of an atom having the degree ${k = 0}$, as
\begin{subequations}
    \begin{align}
        P(s) &= p_0 (1 - p_0)^{s - 1}
        \\
        &= 
        \text{e}^{-n V_s} 
        (1 - \text{e}^{-n V_s})^{s - 1}
        \\
        &\propto \text{e}^{-cs},
    \end{align}
    \label{eq:cluster_sizes}
\end{subequations}
with ${c = - \text{ln}(1 - \text{e}^{-n V_s})}$. In Fig.~\ref{fig:cluster_sizes} a comparison between cluster sizes in Monte-Carlo simulations and eqs.~\eqref{eq:cluster_sizes} is shown. The agreement is very good for small densities as almost all clusters in the gas are composed of linear chains. As the density of the gas increases, the probability that at least one atom in the cluster has more than two connections, i.e. $k \geq 3$, increases. While the distribution remains exponential, the probability for larger clusters to exist in the system greatly increases compared to the prediction by eqs.~\eqref{eq:cluster_sizes}.

\section{Epidemic Dynamics on the Network}

It is known that Rydberg systems in the facilitation regime bear close similarities to epidemics \cite{PhysRevLett.119.140401, natcom_Griffiths}. In the following, we will systematically analyze the Rydberg facilitation dynamics on the random network formed by atoms within their respective facilitation shells. For this, we will map the dynamics to the Susceptible-Infected-Susceptible (SIS) epidemic model \cite{WEISS1971261, murray2002mathematical, bailey1975mathematical}. We will (i) disregard the decay of Rydberg atoms into the dead state $\ket{0}$ (parameter ${b=0}$, see Fig.~\ref{fig:intro}a), reducing the dynamics of each atom to a two-level system. Additionally,  we will (ii) neglect off-resonant excitations by setting ${\tau=0}$, meaning excitations can only be created by means of facilitation. We will refer to simulations carried out with these two constraints as the SIS approximation.

One major difference between our Rydberg system in the SIS approximation and a classical SIS system remains with Rydberg blockade. 
 Atoms excited to the Rydberg state do not only facilitate the spread of excitations, they can also block the spreading in adjacent clusters. This will be analyzed more systematically later.

The network structure of cluster of atoms in facilitation distance of each other strongly depends on the temperature of the gas. 
If the RMS average relative velocity $\overline{v}$ is large, such that each excited Rydberg atom meets many ground state atoms during its
lifetime $\gamma^{-1}$, i.e. if in a 3D gas
\begin{eqnarray}
    \overline{v} \gg \gamma \, n^{-1/3},
\end{eqnarray}
any network structure is effectively washed out and the system is  homogeneous. Close to the critical point of the absorbing-state phase transition, the above condition is equivalent to $ \overline{v} \gg \gamma \, r_f$. 
If, on the other hand, the average velocity of atoms is very small, such that during
a facilitation time $\gfac^{-1}$ they do not move out of the facilitation shell, i.e. if
\begin{equation}
    \overline{v} \ll \gfac \, \delta \rfac,
\end{equation}
the atoms form a finitely connected network. We will now discuss these two limits.

\subsection{High temperature limit}

In a high-temperature gas with RMS average relative velocity $\overline{v} \gg \gamma n^{1/3}$,  we can map the system to the Susceptible-Infected-Susceptible (SIS) epidemic model \cite{WEISS1971261, murray2002mathematical, bailey1975mathematical}. 
The SIS model is 
characterized by the infection and recovery rates, $\lambda$ and $\mu$ respectively, which for the Rydberg gas read
\begin{subequations}
    \begin{align}
        \lambda &= n \kappa
        \\
        \mu &= \gamma.
    \end{align}
\end{subequations}
%

The SIS model predicts an active/absorbing phase transition when 
\begin{equation}
    \label{eq:lambda_c_1}
    \lambda_c^{(1)} = \mu
\end{equation}
where excitation spread equals spontaneous decay. This corresponds to the critical density \eqref{eq:n_crit} of the absorbing state phase transition discussed before.

In Fig.~\ref{fig:ryd_response}a, Monte-Carlo simulations of the Rydberg system with the SIS approximation and ${\rho(t=0) = n}$ are shown for the high temperature gas for different values of $n$ and a fixed  facilitation rate $\gfac$. 
\mfl{We note that as shown in Appendix~\ref{appendix} the excitation probabilities following from Monte-Carlo simulations of rate equations and those from full coherent density matrix simulations agree, showing that the rate equation approach remains valid also in the high-temperature limit. }
One recognizes that an active/absorbing-state phase transition occurs for $\lambda=\lambda_c^{(1)}$, with the Rydberg density either exponentially decaying at the time-scale $\mu$ (for  $\lambda < \lambda_c^{(1)}$), or decaying to a steady-state active density (for $\lambda >\lambda_c^{(1)}$). At the critical density (green curve in Fig.~\ref{fig:ryd_response}a) the system should decay with $\rho \sim t^{-1}$ \cite{munoz2010Griffiths}, however, this decay is truncated by an exponential decay due to finite system size.

\subsection{Frozen gas limit}

In the limit of an effectively frozen gas the atoms that can participate at the facilitation process form a network.
The dynamics of an SIS epidemic strongly depend on the structure of this underlying network. For example, in the case of a heterogeneous but \emph{scale-free} network, i.e. ${P(k) \sim k^{-\nu}}$, the absorbing phase can disappear altogether, leaving the system in an endemic phase regardless of the infection rate \cite{chatterjee2009contact, SciRep.6.22506, RevModPhys.87.925}.

\begin{figure}[H]
    \centering
    \includegraphics[width=0.85\columnwidth]{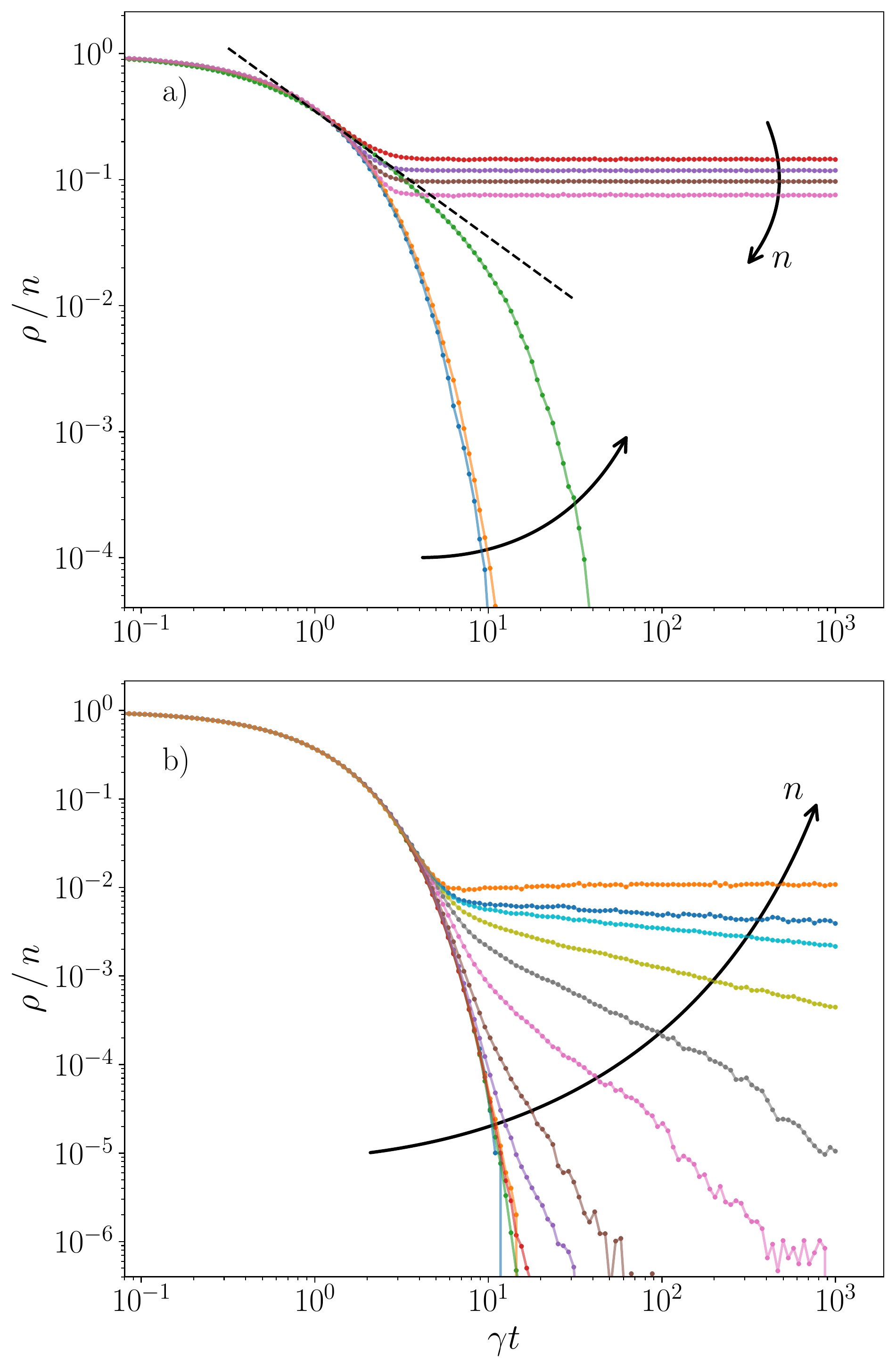}
    \caption{Decay of Rydberg density over time from Monte-Carlo simulations in the SIS approximation (${b=0}$, ${\tau=0}$) with an initial density ${\rho(t=0) = n}$ for the gas at high temperature (a) with ${\hat{v} = 100 \, \rfac \gamma}$ and for the frozen gas (b) with ${\hat{v} = 0 \, \rfac \gamma}$. The black dashed line in (a) is a power-law decay with ${\rho \sim t^{-1}}$, expected at the critical density. The colors show total system densities (increasing from left to right) with ${n=}$ 0.003, 0.03, 0.39, 1.98, 3.97, 5.95, 9.12, 11.90, 14.28, 15.91, 16.66 and 19.94. Figure (a) only shows the lowest 7 densities.
    The critical density is $n_\textrm{crit}=$ 0.39. Between the percolation density $n_\textrm{perc}= $ 15.91 and $n_\textrm{crit}^{(2)}= $ 16.6 the curves feature a decay (see Fig.~\ref{fig:phase-diagram}).
    }
    \label{fig:ryd_response}
\end{figure}

For the case of a heterogeneous ER network, which describes the frozen gas of atoms, an active phase can only occur if the network is above the percolation threshold (i.e. ${\langle k \rangle > 1}$). However, for a finitely connected (but percolating) ER network, the threshold for the active phase is modified since activity occurs in localized regions and thus the effective infection rate is reduced. One finds \cite{munoz2010Griffiths}
\begin{equation}
    \lambda_c^{(2)} = \mu \frac{\langle k\rangle}{\langle k\rangle -1},
\end{equation}

with $\langle k\rangle$ being the average degree of the network. For ${\langle k \rangle \to \infty}$ the threshold given by eq.~\eqref{eq:lambda_c_1} is recovered. For a fixed facilitation rate and facilitation volume this threshold can be expressed in terms of a critical density of atoms using $\langle k\rangle = n V_s$

\begin{align}
    \nonumber
    n_\textrm{crit}^{(2)} &= \frac{1}{V_s} + \frac{\gamma}{\kappa}
    \\
    &\equiv n_\textrm{perc} + n_\textrm{crit}^{(1)}.
\end{align}

If the network is below the percolation threshold, the finite size of clusters truncates the spread of activity through the system. Therefore, the network cannot support an active phase, and instead, a Griffiths phase emerges above the critical infection rate $\lambda_c^{(1)}$ \cite{munoz2010Griffiths}. One of the most distinguishing characteristics of a Griffiths phase is the presence of rare regions with above average activity which lead to a slow, algebraic decay of excitations \cite{PhysRevLett.23.17}.

\begin{figure}[H]
    \centering
    \includegraphics[width=0.8\columnwidth]{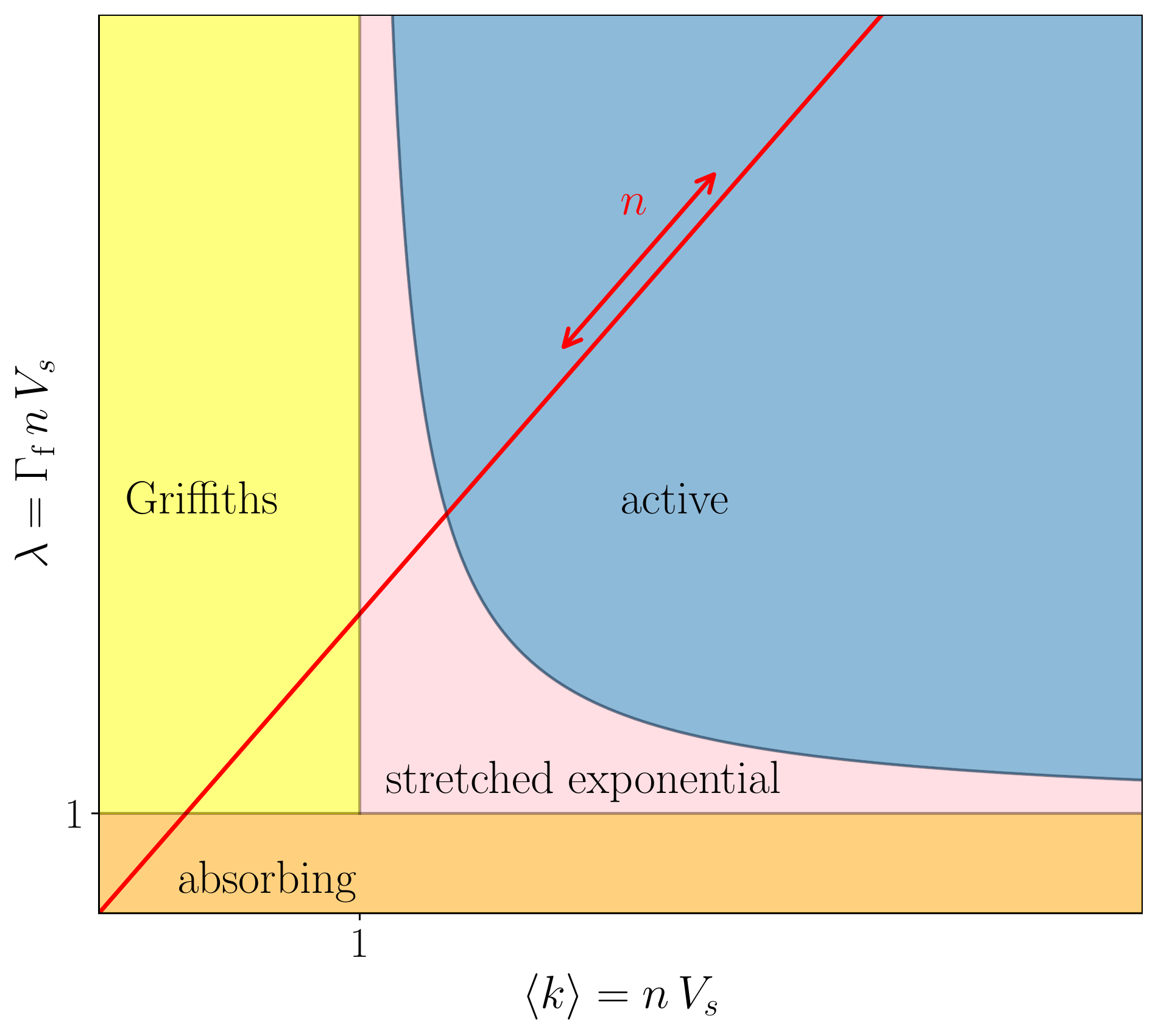}
    \caption{Schematic phase diagram of Rydberg facilitation of a frozen gas with the percolation threshold given at ${\langle k \rangle = 1}$. Increasing the density $n$ of the gas one moves along the red line crossing from an absorbing into a Griffiths phase at $n_\textrm{crit}$, and subsequently into a phase with streched exponential decay at the percolation threshold $n_\textrm{perc}$ and eventually into the active phase at $n_\textrm{crit}^{(2)}$. Time has been rescaled such that ${\mu = 1}$.}
    \label{fig:phase-diagram}
\end{figure}

In the non-percolating network (i.e. ${\langle k \rangle < 1}$), for ${\lambda \leq \mu}$ decay dominates, leading to very short times until all activity disappears in clusters as excitations cannot sustain themselves. If however ${\lambda > \lambda_c^{(1)}= \mu}$, the time until activity disappears in clusters increases exponentially with cluster size $s$ and is given by \cite{10.2307/3215641}
\begin{align}
    \label{eq:cluster_lifetime}
    \tau(s)
    \propto 
    \sqrt{\frac{2 \pi}{s}}
    \frac{\lambda}{(\lambda - 1)^2}
    \text{exp} 
    \Big\{
        s \Big(\text{ln}(\lambda) - 1 + \frac{1}{\lambda}\Big)
    \Big\}.
\end{align}
In the following we will refer to $\tau(s)$ as the extinction time of activity in a cluster. As a result of the convolution of exponentially rare cluster sizes ${P(s) \sim \text{e}^{-cs}}$ and a cluster lifetime increasing exponentially with cluster size ${\tau(s) = \text{e}^{as}}$, the activity in the Griffiths phase decays with a power law:
\begin{align}
    \label{eq:rho_integral}
    \rho(t) = \int \text{d}s \: s P(s) \, \text{e}^{-t / \tau(s)}.
\end{align}
Using eqs.~\eqref{eq:cluster_sizes} and \eqref{eq:cluster_lifetime}, the integral in \eqref{eq:rho_integral} can be approximated with Laplace's method and results in an algebraic decay 
\begin{align}
    \rho \sim t^{-c/a},
\end{align}
with the coefficient $a$ given by \eqref{eq:cluster_lifetime} as ${a = \text{ln}(\lambda) - 1 + \frac{1}{\lambda}}$. 
If the network is above the percolation threshold, i.e. if $\langle k\rangle \ge 1$, but the driving strength is
below the critical value for the active phase $\lambda_c^{(2)}$ the 
decay of activity is expected to follow a stretched exponential. 
A qualitative phase diagram of the facilitation dynamics in the frozen Rydberg gas is shown in Fig.~\ref{fig:phase-diagram}.

Fig.~\ref{fig:ryd_response}b shows the results of Monte Carlo simulations for a frozen gas in the SIS approximation for the same parameters and color code as in the high-temperature case of Fig.~\ref{fig:ryd_response}a. For ${n < n_\textrm{crit}}$ all initial excitations decay exponentially (curves 1 and 2 from left to right), corresponding to the absorbing phase. The behavior changes at and above the critical point but below the percolation threshold ${n_\textrm{crit}\le n < n_\textrm{perc}}$  (curves 3-7). Here, the system is in an extended Griffiths phase with a power-law decay with varying exponents. Above the percolation threshold but below the threshold
of the active phase $n_\textrm{perc} \le n < n_\textrm{crit}^{(2)}$ the decay is expected to become a streched exponential \cite{munoz2010Griffiths}, which we cannot resolve however in our simulations due to the very long time scale of this decay. Finally, for ${n\ge n_\textrm{crit}^{(2)}}$ the system enters the active phase where excitations simply decay to a steady state. 

\begin{figure}[H]
    \centering
    \includegraphics[width=0.8\columnwidth]{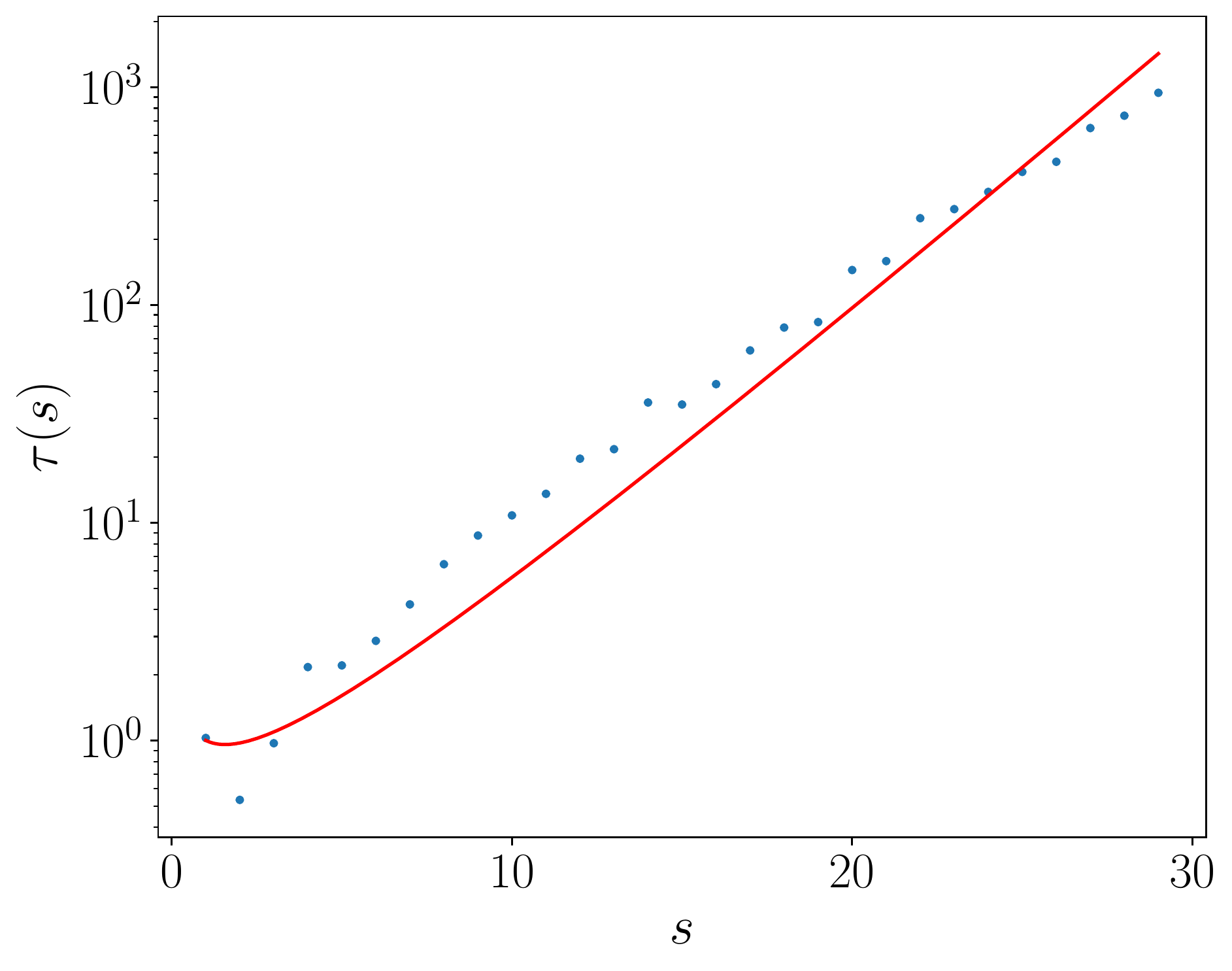}
    \caption{
     Extinction times for activity in clusters in Monte-Carlo simulations of 1D lattice chains of length $s$ (blue dots) and prediction by \cite{10.2307/3215641} (red line).}
     \label{fig:Griffiths_exponents}
\end{figure}

In the following, we want to give a quantitative estimate for the power law decay coefficient in the Griffiths phase based on the SIS model and compare them with those from the Monte-Carlo simulations. In contrast to the standard SIS model, a Rydberg system features Rydberg blockade and facilitated de-excitation, making it unclear if analytic predictions from an SIS model would be accurate. To check this, we compare the extinction time of activity in clusters in a linear excitation chain, given by eq.~\eqref{eq:cluster_lifetime}, using the spreading rate ${\lambda = \gfac V_s \cdot 1 \rfac^{-3}}$, with Monte-Carlo simulations of the SIS approximation in Fig.~\ref{fig:Griffiths_exponents}. For this, we simulate a 1D cluster of length $s$ where each atom is initially in the Rydberg state and measure the average time until all atoms are decayed. Here, we assume the above mentioned SIS approximation (no decay to $\ket{0}$ and no off-resonant excitations). One recognizes that eq.~\eqref{eq:cluster_lifetime} gives a good approximation of the extinction time.

\begin{figure}[H]
    \centering
    \includegraphics[width=0.9\columnwidth]{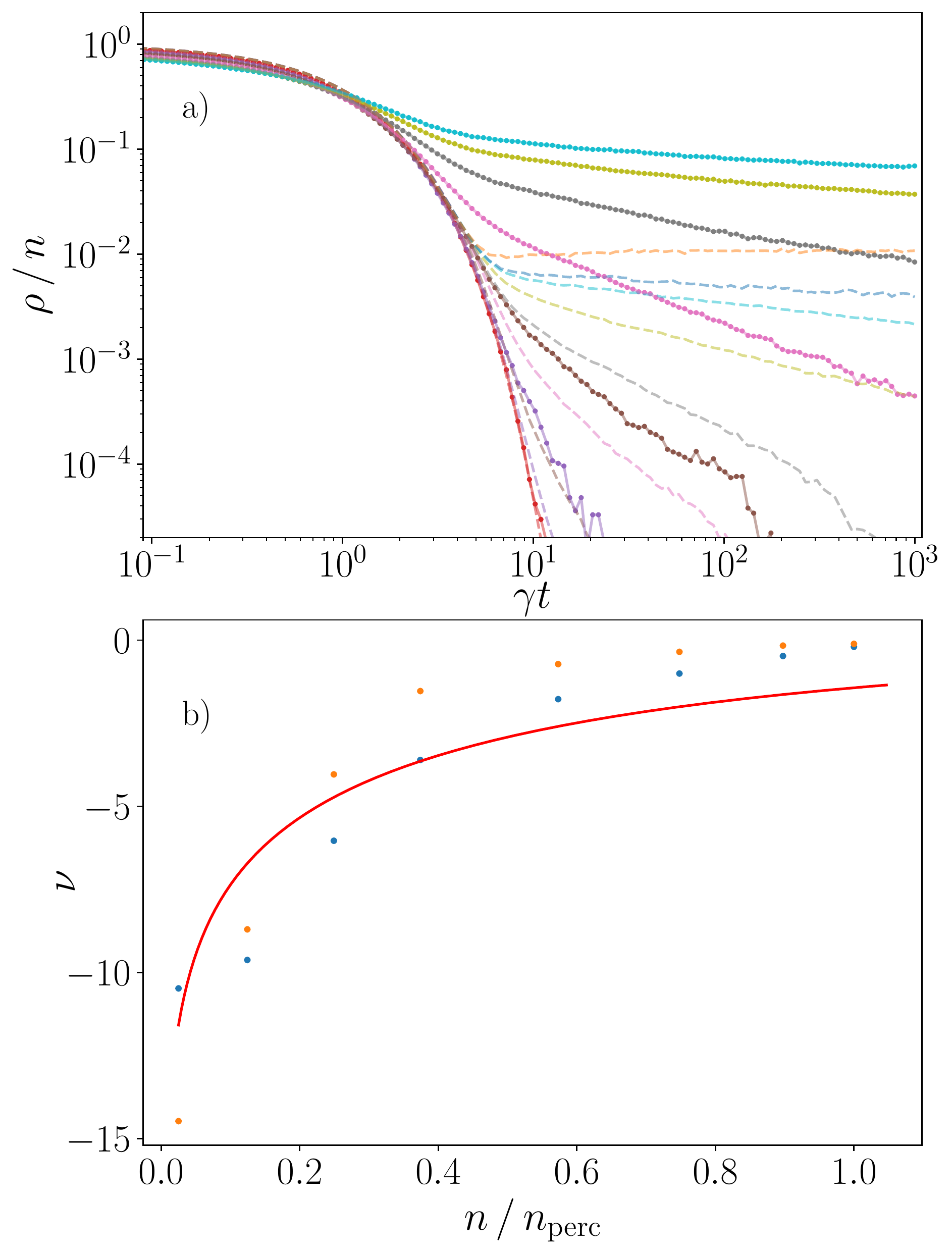}
    \caption{
    a) Decay of Rydberg density over time from Monte-Carlo simulations in the SIS approximation for the frozen gas without Rydberg blockade (full lines) compared to the results from Fig.~\ref{fig:ryd_response}b (dashed lines). The colors show total system densities (increasing from left to right) with n = 0.003, 0.03, 0.39, 1.98,
3.97, 5.95, 9.12, 11.90, 14.28, 15.91, 16.66 and 19.94. 
    b) 
    Power law decay exponent ${\nu = -c/a}$ of Rydberg density over time fitted from frozen gas Monte-Carlo simulations in the SIS approximation from Fig.~\ref{fig:compare_ryd_blockade}a with Rydberg blockade (blue dots), without Rydberg blockade (orange dots), and from the analytical approximation geiven by eq.~\eqref{eq:Griffiths_exponent} (red line).}
    \label{fig:compare_ryd_blockade}
\end{figure}

Using \eqref{eq:cluster_sizes} and \eqref{eq:cluster_lifetime} we can approximate the power-law exponent $\nu$ in the Griffiths phase dependent on the density and internal rates. We receive
\begin{equation}
    \label{eq:Griffiths_exponent}
    \nu \equiv -\frac{c}{a} = -\frac{\ln\left(1-e^{-n V_s}\right)}{\ln(\lambda) -1 +\lambda^{-1}},
\end{equation}
with $\lambda = 4\pi \gfac \frac{\delta \rfac}{\rfac}$. The comparison with exponents fitted from the power law decay of Rydberg density in Monte-Carlo simulations of the frozen gas under the SIS approximation (seen in Fig.~\ref{fig:ryd_response}) can be seen in Fig.~\ref{fig:compare_ryd_blockade}b. Our very rough approximation of the Griffth-phase decay exponents qualitatively fits with Monte-Carlo data.

A fundamental difference between Rydberg facilitation and classical SIS activity spreading is Rydberg blockade. Considering the frozen gas limit, two effects arise from Rydberg blockade: first, if an atom is surrounded by two Rydberg atoms in the facilitation distance, i.e. the atom is in the middle of a cluster, then it cannot be facilitated, as it receives twice the dipole shift and is pushed out of resonance again. If this atom decays or is in the ground state at the beginning, it cannot be excited resulting in a hole splitting the cluster \cite{letscher2017bistability}. Additionally, Rydberg atoms can block excitations from spreading through adjacent clusters. However, neither of these effects change the actual structure of the network, instead they effectively retard the timescale at which excitations spread. For a quantitative comparison we simulate the Rydberg gas and compare the decay of excitations in the SIS approximation with and without Rydberg blockade (Fig.~\ref{fig:compare_ryd_blockade}).

As blockade allows fewer Rydberg atoms to be present in the system, the steady state Rydberg density of the active phase, and therefore the density at which the power law decay of the Griffiths phase begins is much lower. However, as seen in Fig.~\ref{fig:compare_ryd_blockade}, the exponents of the power-law decay in the Griffiths phase show no qualitative change depending on the presence of Rydberg blockade in the system.

\section{Conclusion}

We studied the facilitation dynamics of Rydberg excitations in an ultra-cold gas of atoms. While in the homogeneous limit, the system is expected to show a phase transition between an absorbing phase and an active phase, and -- in the presence of an additional loss channel from the Rydberg state -- self-organized criticality (SOC). \daniel{However,} experiments with a gas of trapped $^{87}$Rb atoms at low temperatures show signs of scale invariant dynamics in an extended  parameter regime, \daniel{which is a} signature of a Griffiths phase replacing the critical point of the absorbing-state phase transition. 

To understand 
the emergence of \daniel{scale invariance in the experiment} \mfl{in an extended parameter regime}, we numerically
simulated the many-body Rydberg gas in the facilitation regime
through the use of Monte-Carlo simulations in the classical rate-equation approximation. We showed that the latter is well justified for the large dephasing characteristic for the experiment even for a high-temperature gas. Since a Griffiths phase originates from heterogeneity in the system, we numerically and theoretically analyzed two limiting cases: (i) a high-temperature gas and
(ii) a frozen gas. 
While in the high-temperature limit, a homogeneous
mean-field behavior is recovered, with a clear absorbing-state
phase transition and SOC, \mfl{the} facilitation dynamics in a low-temperature or frozen gas is governed by the presence of a network structure of atoms that can participate in the excitation spread. 
Numerical simulations show characteristic power-law decay of Rydberg excitations in time if off-resonant excitations and atom losses are neglected. 

We have shown that in the frozen gas the spread of excitations is constrained to a network resembling a random Erdős–Rényi (ER) graph.
Increasing the density of atoms the ER network has a percolation transition from a fragmented phase, in which the maximum cluster size of connected atoms remains finite, to a phase where the size of the largest cluster scales with the size of the system. 
A theoretical explanation of the Rydberg facilitation dynamics observed in Monte-Carlo simulations can then be given by mapping  to a susceptible-infected-susceptible (SIS) epidemic model on such 
an ER graph taking into account the effects of Rydberg blockade, which truncates the maximum Rydberg excitation density. 
An active phase of self-sustained Rydberg excitations is only possible above the percolation threshold.
Below this threshold, an extended Griffiths phase emerges in the place of the (for homogeneous systems) expected absorbing-state phase transition. We showed that the modified SIS model quantitatively explains the observed power-law decay exponents as well as the overall dynamics of the Rydberg density. 

While the limits of a high-temperature and a frozen gas are well captured with our model, it does not yet allow the study of the crossover
between the two regimes. To this end, the Rydberg facilitation process needs to be mapped to a dynamical network, which is beyond the scope of the present work and will be the subject of future work. 
Furthermore, in order to quantitatively understand the power-law exponents in the number distribution of Rydberg atoms in a given time interval observed in the experiment, it is necessary to extend the microscopic simulations to much larger system sizes matching those used in the experiments. To this end different approaches, e.g. using machine-learning algorithms might be useful \cite{ohler2022towards}. Finally, the interplay between coherent quantum dynamics and dynamical network structures in
Rydberg facilitation under conditions where dephasing is much less dominant could give rise to very different dynamics \cite{mattioli2015classical,PhysRevLett.116.245701}. The latter requires, however, the development of new microscopic simulation techniques capable of incorporating quantum coherences in 3D Rydberg gases, at least in an approximate way \cite{PhysRevResearch.4.043136}.

\ 

\subsection*{Acknowledgement}

The authors thank Johannes Otterbach for fruitful discussions. 
Financial support from the DFG through SFB TR 185, project number
277625399, is gratefully acknowledged.

\subsection*{Authors contributions}
DB and MF developed the theoretical models, DB performed all numerical simulations and developed the mapping to the ER network \mfl{with support by SO}. MF, TN, and HO conceived the project. JB, PM, and DB performed the experiments guided by TN and HO. DB and MF wrote the initial version of the manuscript with support by JB. All authors discussed the results and contributed to the writing of the manuscript.

\bibliography{references}

\newpage
\appendix

\section{Effects of relative motion between atoms}
\label{appendix}

The rate equation approximation \daniel{used for the Monte-Carlo simulations (e.g. eq.~\ref{eq:gamma_fac})} is valid as long as the population dynamics are slow compared to the dephasing rate. In a frozen gas, the relevant time scales 
are solely determined by the internal dynamics of an individual atom for a given (fixed) configuration of Rydberg atoms in its vicinity. If however, the gas of atoms has a finite temperature, a ground state atom can fly in and out of the facilitation volume $V_s$ of a Rydberg atom, which can amount to a fast sweep of the detuning of the ground state atom. Thus there is an additional time scale given by the crossing time $\sim \delta r_f/v$.

\begin{figure}[H]
    \centering
    \includegraphics[width=0.9\columnwidth]{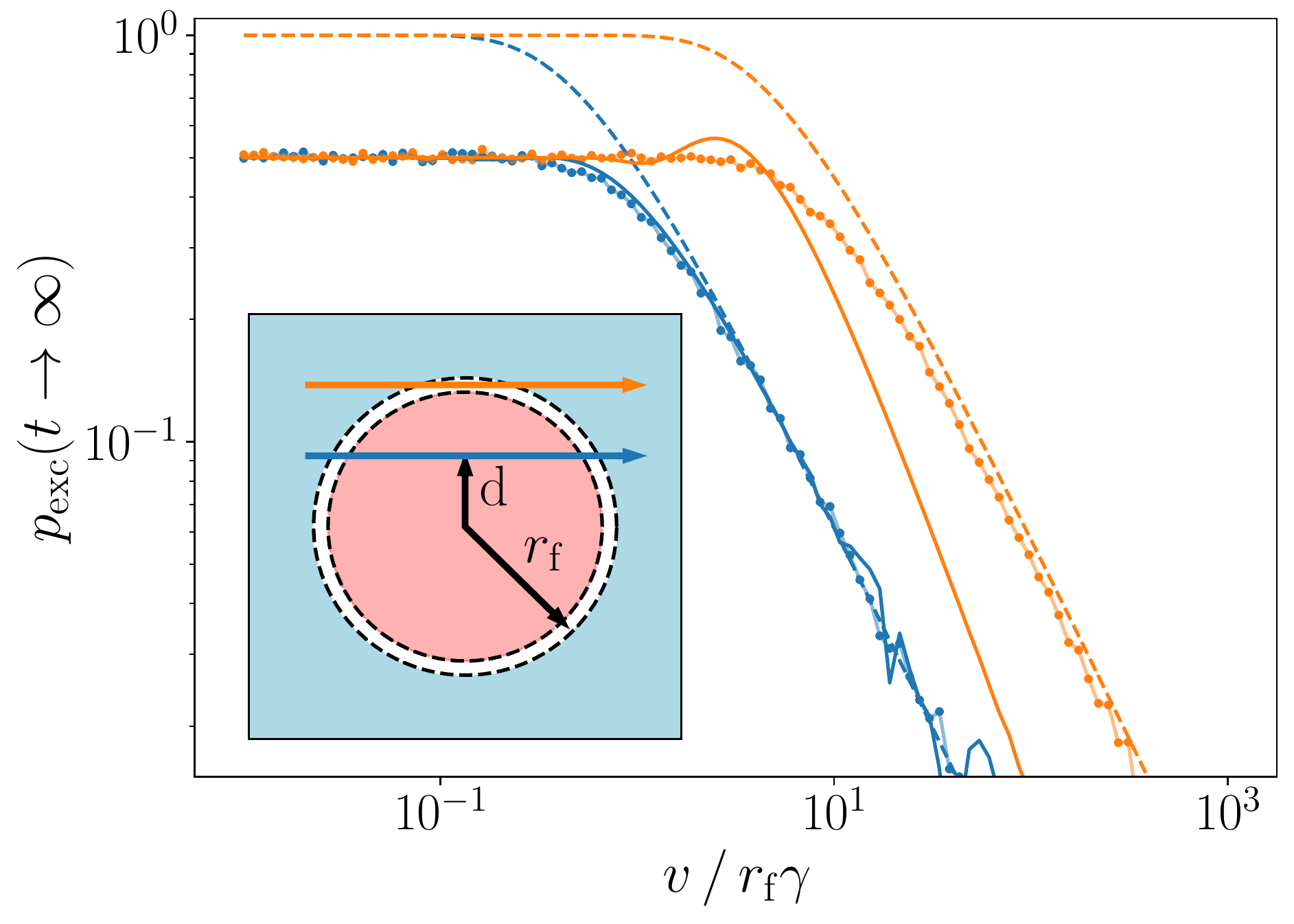}
    \caption{Velocity dependent excitation probability of a ground state atom flying past a Rydberg atom for the impact parameter ${d=0.5 \rfac}$ (blue) and ${d = \rfac}$ (orange) (see inset) calculated from Monte-Carlo simulations with fixed time step ${dt = 1/400 \, \gamma}$ (dots), full numeric density matrix simulation (solid lines), and analytical Landau-Zener formula eq.\eqref{eq:Landau-Zener} (dashed lines). Full solution and rate equation approximation only differ for the case of "grazing incidence" ($d=\rfac$).}
    \label{fig:rate-eqs-vs-full}
\end{figure}

To analyze the effects of atomic motion onto the facilitation process we consider the two-body problem of a ground state atom moving with velocity $v$ and with the impact parameter $d$ relative to a Rydberg atom (see inset of Fig.~\ref{fig:rate-eqs-vs-full}). For ${d > \rfac}$ the ground state atom is not shifted into resonance and no facilitation occurs. For ${d \leq \rfac}$ one has to distinguish two cases depending on the impact parameter: (i) ${d < \rfac}$ the ground state atom flies through the facilitation shell twice (blue case in Fig.~\ref{fig:rate-eqs-vs-full}), and (ii) ${d \approx \rfac}$ the ground state atom grazes the facilitation shell and is only briefly shifted into resonance (orange case in Fig.~\ref{fig:rate-eqs-vs-full}). 

In case (i), \daniel{using the excitation rate from eq.~\eqref{eq:gamma_fac} as $\Gamma_\uparrow(t)$,} we find the excitation probability after a single pass of the ground state atom through the facilitation shell as
\begin{eqnarray}
    p_\text{exc} &=& 1-\exp\left\{-\int_{t_i}^{t_f}\!\!\! \text{d}t\, \Gamma_\uparrow(t)\right\}\nonumber\\
    &=& 1-\exp\left\{-2 \Omega^2  \int_{t_i}^{t_f}\!\!\! \text{d}t\, \frac{\gamma_\perp}{\Delta(t)^2+\gamma_\perp^2}\right\}.
\end{eqnarray}
\daniel{Note that this expression assumes a short passage time through the facilitation shell, so that the facilitated de-excitation process can be ignored. For longer passage times the excitation probability approachs the steady state value of ${1/2}$, as can be seen Fig.~\ref{fig:rate-eqs-vs-full}.}

Linearizing the time dependent detuning $\Delta(t)$ for times $t_i< t < t_f$, while passing through the facilitation shell, we receive $\Delta(t) \approx \dot\Delta \times  (t-t_0)$ yielding
\begin{eqnarray}
    p_\text{exc} &=& 1 - \exp\left\{-\frac{2\Omega^2}{\dot\Delta}\int_{\Delta_i}^{\Delta_f}\!\! \! \! d\Delta \, \frac{\gamma_\perp}{\Delta^2+\gamma_\perp^2}\right\}\nonumber\\
    & \approx& 1 - \exp\left\{-2\pi \frac{\Omega^2}{\dot\Delta}\right\},
    \label{eq:Landau-Zener}
\end{eqnarray}

where we have assumed that $\vert \Delta_{i,f}\vert =\vert\Delta(t_{i,f})\vert \gg \gamma_\perp$, which is exactly the same expression as given by the Landau-Zener formula.

If $p_\text{exc}$ is small, the asymptotic excitation probability after two passages is just
${p_\text{exc} \approx 1-\exp\{-4\pi \Omega^2/\dot \Delta\}}$. From this discussion we expect the rate equations to accurately describe the facilitation
process even for large atom velocities as long as the impact parameter $d$ is different from $r_f\pm \delta r_f$. 

In case (ii) however,
i.e. for "grazing incidence", the Landau-Zener formula is no longer valid and there could be a difference between the rate-equation approximation and the solution of the full two-particle density matrix equations. This is indeed the case, as can be seen from Fig.~\ref{fig:rate-eqs-vs-full}, where we have plotted  the asymptotic excitation probability of the ground state atom as function of relative velocity and impact parameter both from a simulation of the full density-matrix equations (dashed lines), the analytic Landau Zener formula (solid line), and by a Monte-Carlo simulation of the rate equation in the large-dephasing limit with time step ${dt = 1/400 \, \gamma}$ (dots).  One recognizes perfect agreement except for large relative velocities and impact parameters close to the facilitation radius ${d\approx \rfac}$, where the rate equations predict up to an order of magnitude higher excitation probabilities than the full simulation. Since ${\delta \rfac\ll \rfac}$ the contribution of these "grazing-incidence" cases is negligibly small, allowing us to accurately describe high gas temperatures with a fixed time step Monte-Carlo algorithm.

\daniel{Furthermore, at high temperatures (as can be seen from Fig.~\ref{fig:rate-eqs-vs-full}) the excitation probability 
above the Landau-Zener velocity ${v_\textrm{LZ}= 2\pi^2 \Omega^2 \rfac / (3\Delta)}$ indeed quickly drops and scales as $1/v$,  the number of ground state atoms seen by a moving Rydberg atom in a given time increases linearly with its velocity $v$, too. This compensates the former effect, and thus does not lead to an emerging heterogeneity in phase space as argued in \cite{natcom_Griffiths}, as long as the Rydberg atom does not move out of the gas sample.}

\end{document}